# Search Process and Probabilistic Bifix Approach

Dragana Bajic
Faculty of Technical Sciences
University of Novi Sad
Trg Dositeja Obradovica 6
21000 Novi Sad,
Serbia&Montenegro
Email: lmcdra@eunet.yu

Cedomir Stefanovic
Faculty of Technical Sciences
University of Novi Sad
Trg Dositeja Obradovica 6
21000 Novi Sad
Serbia&Montenegro
Email: cex@uns.ns.ac.yu

Dejan Vukobratovic
Faculty of Technical Sciences
University of Novi Sad
Trg Dositeja Obradovica 6
21000 Novi Sad
Serbia&Montenegro
Email: dejanv@uns.ns.ac.yu

*Abstract* — **An analytical approach to a search process is a mathematical prerequisite for digital synchronization acquisition analysis and optimization. A search is performed for an arbitrary set of sequences within random but not equiprobable *L*-ary data. This paper derives in detail an expression for probability distribution function, from which other statistical parameters - expected value and variance – can be obtained. The probabilistic nature of (cross-) bifix indicators is shown and application examples are outlined, ranging beyond the usual telecommunication field.**

## I. Introduction

An objective of the search process is to find one of the $M$ predefined sequences of length $N$ while performing the tests along a stream of random $L$-ary data. The process itself is simple, but the statistics describing it were not, so the first ones were obtained by simulation study.

A pioneering analytical approach [1] has introduced an ingenious feature of "bifix", named by Prof. J. Massey. A bifix is a subsequence that is both a prefix and a suffix of a longer sequence. It enabled the evaluation of expected value of number of tests taken to observe the first appearance of a single sequence in equiprobable data. Equivalent but independent study, at binary level only, was performed by mathematicians and summarized in [2].

Further research, by this author, added the probability distribution function and variance to already known expected value [3]. The results were later extended to the case of set of $M$ sequences [4]. Since the applications that have followed [5, 6 and 7] were related to digital transmission assuming scrambled data, the derived formulae at equiprobable level were considered complete. A motivation to switch to non-equiprobable case was initiated by a suggestion of anonymous reviewer of [6].

The next section introduces the basic elements necessary for derivation. Within the third section the statistical parameters for $M=1$ case are derived, extended to $M>1$ case within the section IV. The possible application and topics for further research are outlined within the concluding remarks.

## II. Cross-Bifix Spectrum and Tail Vectors

An infinite **stream** of data consists of random $L$-ary symbols $x: x \in \{X_1, X_2, \cdots, X_L\}$, with $\Pr\{x=X_i\}=p_i$, $i=1,\cdots,L$, and $\sum_{i=1}^{L} p_i = 1$. Therefore, the symbols are **not** equiprobable.

Starting at the random point, a search is performed, stopping at the moment when the first $N$-symbols long **sequence**, out of the set of $M$ predefined sequences, is observed.

From the set of sequences a "cross-bifix spectrum" and a set of $M$ "tail" vectors can be derived.

A cross-bifix is a subsequence of length $n$, $n \leq N$, that is a suffix of one sequence and a prefix of another. Its indicator is denoted $h_{ij}^{(n)}$, where subscripts $i$ and $j$ denote respectively sequences which suffix and prefix are observed, and superscript $(n)$ denote the cross-bifix length. The default values are $h_{ii}^{(N)}=1$ (a sequence is equal to itself), $h_{ij}^{(N)}=0$, $i \neq j$ (no two sequences out of $M$ ones are the same) and $h_{ij}^{(0)}=1$ (for the reasons that would be clear after the next section), $i,j \leq M$.

Cross-bifix spectrum consists of $N+1$ cross-bifix $M \times M$ matrices, shown for an example **010** and **100** ($M=2$, $L=2$, $N=3$):

$$\mathbf{h}^{(0)} = \begin{bmatrix} 1 & 1 \\ 1 & 1 \end{bmatrix}, \quad \mathbf{h}^{(1)} = \begin{bmatrix} 1 & 0 \\ 1 & 0 \end{bmatrix}, \quad \mathbf{h}^{(2)} = \begin{bmatrix} 0 & 1 \\ 0 & 0 \end{bmatrix}, \quad \mathbf{h}^{(N)} = \begin{bmatrix} 1 & 0 \\ 0 & 1 \end{bmatrix}.$$

Last $n$, $n \leq N$, symbols of a sequence no. $i$ form its

"tail" – a suffix. Its probability $r_i^{(n)}$ is product of a probabilities of this last $n$ symbols. A tail vector consists of all tail probabilities with a default $r_i^{(0)}=1$, $i=1,\cdots,M$. For the previous example, if $\Pr\{x=1\}=p_1=p$ and $\Pr\{x=0\}=p_2=q$, two tail vectors are:
$$\mathbf{r}_1 = \begin{bmatrix} 1 & q & p\cdot q & p\cdot q^2 \end{bmatrix}, \quad \mathbf{r}_2 = \begin{bmatrix} 1 & q & q^2 & p\cdot q^2 \end{bmatrix}.$$
The longest tail is equal to the sequence itself.

In case of a single search the bifix spectrum is reduced to $(N+1)$ scalars and there is just one tail vector. The subscripts are not necessary, so $h_{11}^{(n)}=h^{(n)}$ and $r_1^{(n)}=r^{(n)}$, $n=1,\ldots,N$.

Statistical parameters that will be derived within the following sections are:
- $\Pr\{k\}$ - probability that one of the sequences would be found after exactly $k$ tests – a probability distribution function;
- $E\{k\} = T$ – expected duration of a search;
- $E\{k^2\}$ – second moment;
- $\sigma^2 = E\{k^2\} - E^2\{k\}$ – variance.

## III. IN QUEST FOR A SINGLE SEQUENCE

Suppose that a chosen sequence $P$ of length $N$ is found at the test position $k$. The probability of this event, $\Pr\{k\}$, is actually a probability of a set of all $k$-*constrained parent streams* ($k$-CPS) – streams with a property (a constraint) that no $N$ successive symbols match to the observed sequence $P$, except the last $N$ ones. Any stream that contains sequence $P$ prior the position $k$ is considered *unconstrained*.

All $k$-constrained and $k$-unconstrained streams form sets $CS^{(k+N-1)}$ and $US^{(k+N-1)}$ respectively, where superscripts denote the stream length. Since the first $(k-1)$ symbols are arbitrary and the sets are disjunctive, the following may be written:

$$\Pr\{CS^{(k+N-1)} + US^{(k+N-1)}\} = \\ = \Pr\{CS^{(k+N-1)}\} + \Pr\{US^{(k+N-1)}\} = r^{(N)}. \quad (1)$$

A search performed along a $k$-constrained stream succeeds for the first time at the $k^{th}$ trial, so $\Pr\{CS^{(k+N-1)}\} = \Pr\{k\}$ and Eq. (1) can be rearranged as:

$$\Pr\{k\} = r^{(N)} - \Pr\{US^{(k+N-1)}\}. \quad (2)$$

A set $US^{(N+k-1)}$ can be further decomposed into $k-1$ disjunctive subsets, as shown on Fig. 1:

$$US^{(k+N-1)} = \sum_{f=1}^{k-1} US_f^{(k+N-1)} \quad (3)$$

where index "$f$" denotes the position of the first occurrence of sequence $P$, $f<k$. Each stream from a subset $US_f^{(k+N-1)}$ consists of an $f$-constrained stream ($P$ has appeared for the first time at the $f^{th}$ position!), arbitrary stream of length $k$-$N$-$f$ ($L^{k-N-f}$ possibilities) and a sequence $P$ (Fig. 1a). The probability of this subset is:

$$\Pr\{US_f^{(k+N-1)}\} = \\ = \Pr\{CS^{(f+N-1)}\}\cdot\Pr\{\text{all }L^{k-N-f}\text{ streams}\}\cdot r^{(N)} = \\ = \Pr\{f\}\cdot r^{(N)}, \quad f \leq k-N \quad (4)$$

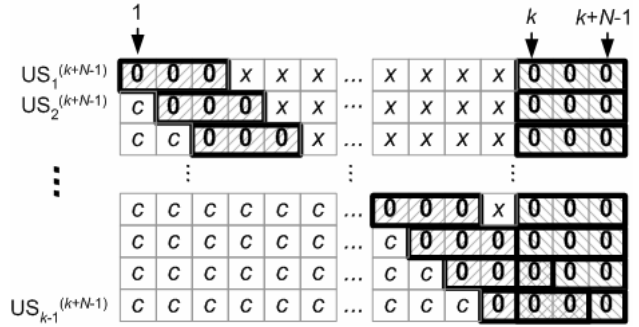

Figure 1: Decomposition of unconstrained stream

For $f > k-N$, an $f$-constrained stream and sequence $P$ overlap, so Eq. (4) does not hold. A $k$-unconstrained stream would exist only if the overlapping symbols enable it, i.e. if the bifix of length $N-(k-f)$ exists. Then a tail of length $k-f$ attached to the bifix (and to the $f$-constrained stream) forms a sequence $P$ at the end of the stream (Fig. 1b). The probability of the corresponding $k$-unconstrained subset is:

$$\Pr\{US_f^{(k+N-1)}\} = \Pr\{f\}\cdot h^{(N-(k-f))}\cdot r^{(k-f)}, \\ f > k-N \quad (5)$$

Summation of Eqs. (4) and (5) according to Eq. (3) and substituting $m=k-f$ yields:

$$\Pr\{US^{(k+N-1)}\} = \begin{cases} \sum_{m=1}^{k-1} h^{(N-m)}\cdot r^{(m)}\cdot \Pr\{k-m\}, \\ \qquad k \leq N+1 \\ \sum_{m=1}^{N} h^{(N-m)}\cdot r^{(m)}\cdot \Pr\{k-m\} + \\ + r^{(N)}\cdot \sum_{m=N+1}^{k-1} \Pr\{k-m\}, \\ \qquad k > N+1 \end{cases} \quad (6)$$

Eq. (2) holds for any $k$, so it holds for $k-1$ as well. If $k-1>N+1$ it may be written:

$$r^{(N)} = \Pr\{k-1\} + \Pr\{US^{(k-1+N-1)}\} =$$
$$= \Pr\{k-1\} + \sum_{l=1}^{N} h^{(N-l)} \cdot r^{(l)} \cdot \Pr\{k-1-l\} +$$
$$+ r^{(N)} \cdot \sum_{l=N+1}^{k-2} \Pr\{k-1-l\} \quad (7)$$

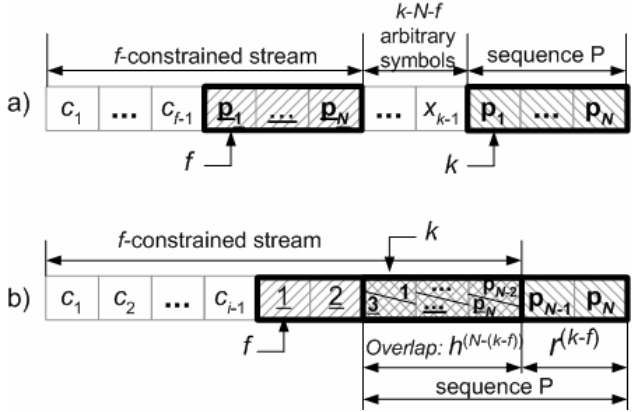

Fig. 2: An example of $k$-unconstrained stream
a) $f \le k-N$;
b) $f > k-N$;

Inserting Eqs. (6) (for $k > N+1$) and (7) into Eq. (2) and substituting $l = m-1$ $\Pr\{k\}$ is evaluated:

$$\Pr\{k\} = \Pr\{k-1\} + \sum_{m=2}^{N+1} h^{(N-m+1)} \cdot r^{(m-1)} \cdot \Pr\{k-m\} +$$
$$+ r^{(N)} \cdot \sum_{m=N+2}^{k-1} \Pr\{k-m\} - \sum_{m=1}^{N} h^{(N-m)} \cdot r^{(m)} \cdot \Pr\{k-m\} -$$
$$- r^{(N)} \cdot \sum_{m=N+1}^{k-1} \Pr\{k-m\} =$$
$$= h^{(N)} \cdot r^{(0)} \cdot \Pr\{k-1\} + h^{(0)} \cdot r^{(N)} \cdot \Pr\{k-N-1\} -$$
$$- h^{(N-1)} \cdot r^{(1)} \cdot \Pr\{k-1\} +$$
$$\sum_{m=2}^{N} \left( h^{(N-m+1)} \cdot r^{(m-1)} - h^{(N-m)} \cdot r^{(m)} \right) \cdot \Pr\{k-m\} -$$
$$- r^{(N)} \cdot \Pr\{k-N-1\} =$$
$$= \sum_{m=1}^{N} \left( h^{(N-m+1)} \cdot r^{(m-1)} - h^{(N-m)} \cdot r^{(m)} \right) \cdot \Pr\{k-m\} \quad (8)$$

Having in mind the initial probability (for $k=1$) and the first couple of terms when $k \le N+1$, the final expression for probability distribution function in a case of a search for $M=1$ sequence is:

$$\Pr\{k\} = \begin{cases} \Pr\{\text{sequence}\} = r^{(N)}, & k=1 \\ \sum_{m=1}^{\min(N,k-1)} \left( h^{(N-m+1)} \cdot r^{(m-1)} - h^{(N-m)} \cdot r^{(m)} \right) \cdot \Pr\{k-m\}, & k>1 \end{cases} \quad (9)$$

From Eq. (8) it can be seen that the indicator $h^{(0)}$ and the probability $r^{(0)}$ of length zero are introduced formally and for this reason the finite summation (9) is evaluated in a symmetrical form.

Starting from the infinite summations:

$$E\{k\} = \sum_{k=1}^{\infty} k \cdot \Pr\{k\}, \qquad E\{k^2\} = \sum_{k=1}^{\infty} k^2 \cdot \Pr\{k\} \quad (10)$$

the required first and second moments are evaluated in terms of finite sums:

$$E\{k\} = T = 1 - N + \sum_{m=1}^{N} h^{(N-m+1)} \cdot \frac{r^{(m-1)}}{r^{(N)}} \quad (11)$$

$$E\{k^2\} = 2 \cdot T^2 - T + N - N^2 + 2 \cdot \sum_{m=1}^{N} m \cdot h^{(N-m+1)} \cdot \frac{r^{(m-1)}}{r^{(N)}} \quad (12)$$

$$s^2 = (T-N) \cdot (T+N-1) + 2 \cdot \sum_{m=1}^{N} m \cdot h^{(N-m+1)} \cdot \frac{r^{(m-1)}}{r^{(N)}} \quad (13)$$

## IV. SET OF $M$ SEQUENCES

If the search is performed until any of $M$ predefined sequences is observed, $M$ sets $k$-constrained streams $CS^{(k+N-1),j}$ are defined, one for *each* one of the sequences $P_j$. The corresponding probabilities are $\Pr^{(j)}\{k\}$, $j=1,...,M$. A related set of $k$-unconstrained streams $US^{(k+N-1),j}$ for sequence $S_j$, $j=1,...,M$ can further be decomposed into the subsets where the first occurrence of *any* sequence $S_i$, $i=1,...,M$, is at the position $f$:

$$US^{(k+N-1),j} = \sum_{i=1}^{M} \sum_{f=1}^{k-1} US_{f,i}^{(k+N-1),j}, \quad j = 1,\cdots,M \quad (14)$$

The probability of set $US^{(k+N-1),j}$ can be derived similarly to Eq. (6), with a summation that is a consequence of $M$ possible sequences at the position $f$:

$$\Pr\{US^{(k+N-1),j}\} = \begin{cases} \sum_{i=1}^{M} \sum_{m=1}^{k-1} h_{ij}^{(N-m)} \cdot r_j^{(m)} \cdot \Pr^{(i)}\{k-m\}, \\ \qquad k \le N+1 \\ \sum_{i=1}^{M} \sum_{m=1}^{N} h_{ij}^{(N-m)} \cdot r_j^{(m)} \cdot \Pr^{(i)}\{k-m\} + \\ + r_j^{(N)} \cdot \sum_{i=1}^{M} \sum_{m=N+1}^{k-1} \Pr^{(i)}\{k-m\}, \\ \qquad k > N+1 \end{cases} \quad (15)$$

Applying Eq. (2) to the case of $M$ sequences, the probability that the search stops at the position $k$ is:

$$\Pr\{k\} = \sum_{j=1}^{M} \Pr^{(j)}\{k\} = \sum_{j=1}^{M} \left( r_j^{(N)} - \Pr\{US^{(k+N-1),j}\} \right) \quad (16)$$

Finally, following the same procedure as for Eq. (8) and knowing that the starting probability is the

sum of probabilities of all sequence, the probability distribution function is:

$$\Pr\{k\} = \sum_{i=1}^{M} r_i^{(N)},$$
$$\Pr\{k\} = \sum_{j=1}^{M}\sum_{i=1}^{M}\sum_{m=1}^{\min(N,k-1)} \left(h_{ij}^{(N-m+1)} \cdot r_j^{(m-1)} - h_{ij}^{(N-m)} \cdot r_j^{(m)}\right) \cdot \Pr^{(i)}\{k-m\} \quad (17)$$

The moments are expressed as:

$$E\{k\} = T = 1 - N + [\Pr\{1\}]^{-1} \cdot \sum_{i=1}^{M} S_i \cdot \sum_{j=1}^{M} C_{ij} \quad (18)$$

$$E\{k^2\} = 1 - 2 \cdot N \cdot T - N^2 + [\Pr\{1\}]^{-1} \cdot \sum_{j=1}^{M}\sum_{i=1}^{M}(2 \cdot T_i \cdot C_{ij} + S_i \cdot W_{ij}) \quad (19)$$

where

$$C_{ij} = \sum_{m=1}^{N} r_j^{(m-1)} \cdot h_{ij}^{(N-m+1)}; \quad W_{ij} = \sum_{m=1}^{N} (2m-1) \cdot r_j^{(m-1)} \cdot h_{ij}^{(N-m+1)}. \quad (20)$$

Unfortunately, the evaluations of quantities $S_i$ and $T_i$ is more cumbersome. Corresponding vectors $\mathbf{S}=[S_1, S_2, ..., S_M]$ and $\mathbf{T}=[T_1, T_2, ..., T_M]$ are results of the set of linear equations: $\mathbf{A} \cdot \mathbf{S}^T = \mathbf{0}$, $\sum_{i=1}^{M} S_i = 1$; and $\mathbf{A} \cdot \mathbf{T}^T = \mathbf{B}$, $\sum_{i=1}^{M} T_i = T$; where the coefficients are:

$$A_{ij} = C_{j1} \cdot [r_1^{(N)}]^{-1} - C_{j,i+1} \cdot [r_{i+1}^{(N)}]^{-1}$$
$$B_i = \sum_{j=1}^{M} 0.5 \cdot (W_{j,i+1} \cdot [r_{i+1}^{(N)}]^{-1} - W_{j1} \cdot [r_1^{(N)}]^{-1}) \cdot S_j \quad (21)$$

The Equations (18-21), as well as Equations (11-13) are the result of tedious but straightforward summations which are available upon request.

It is interesting to add that, if sequences are bifix free ($M=1$) or cross-bifix free ($M>1$), i.e. if all the (cross-) bifix indicators are equal to 0 except the default ones, the expected duration of a search depends only upon the sequences' length and the probability that the first test is successful (the probability of sequences themselves):

$$E\{k\} = T = 1 - N + [\Pr\{1\}]^{-1} \quad (22)$$

## V. CONCLUDING REMARKS

The major application of these and previously derived formulae is frame acquisition time analysis and its optimization, especially regarding the framing sequence structure. So far, the cros bifix indicators have been regarded as a mere sequence descriptors. But, the probabilities of unconstrained streams are expressed as a product of another probability, tail probability and a (cross-) bifix indicator. On the other hand, a cross-bifix indicator is an indicator function so it has a binary probabilistic structure. The idea that follows is to *regard these indicators as the probability of overlapping* that is not necessarily 0 or 1 – if we compare a received erroneous sequence with a correct locally generated one, the overlapping with depend upon the amount of noise and interference. The consequence of this important conclusion is the possibility to form a "soft" cross-bifix spectrum, i.e. cross-bifix indicators that would include the impact of channel errors.

Design of optimal set of sequences for various purposes (e.g. sets of sequences for MIMO application) using the features extracted from (soft) cross-bifix spectrum is another line of research. Special attention is devoted to specific transmission systems without the scramblers, to quaternary sequences (because of QPSK) and to distributed sequences [8], due to their good overall properties. Another similar application is the optimization of position of pilot sequence symbols (vs. uniformly distributed pilot symbols) in case of data-aided burst synchronizers.

Complete unrelated application lay within the field of biomedical signals, where exists a necessity to compare a measured time before a "smooth" region of heart-rate signal appears with some reference; and the reference is calculated using the derived expression thus avoiding the unpopular but inevitable usage of "surrogate data".